\documentclass[aps,prb,reprint,floatfix,showpacs,amsmath,amssymb]{revtex4-1}
\usepackage{natbib}
\usepackage{float}
\usepackage[dvips,final]{graphicx}
\usepackage{color}
\usepackage{amsmath}
\usepackage{mathrsfs} % fancy hamiltonian

\usepackage{graphicx, subfigure}
\usepackage{pslatex}
\usepackage{relsize}
\usepackage{bm}
\usepackage{xspace} % nice spacing for math/text symbols

\usepackage{xcolor}
\usepackage{soul}

\def\ba{\begin{eqnarray}}
\def\ea{\end{eqnarray}}
\def\be{\begin{equation}}
\def\ee{\end{equation}}

\bibpunct{[}{]}{,}{n}{}{} % reset to online citations

\begin{document}

\title{Effective Potential for Emergent Majorana Fermions in Superconductor Systems}

\author{A. W. Teixeira}
\email{allwtx@gmail.com}
\affiliation{Departamento de F\'isica, Universidade Federal de Vi\c cosa, 36570-900,
Vi\c cosa, Brazil}

\author{V. L. Carvalho-Santos}
\email{vagson.santos@ufv.br}
\affiliation{Departamento de F\'isica, Universidade Federal de Vi\c cosa, 36570-900,
Vi\c cosa, Brazil}

\author{J. M. Fonseca}
\email{jakson.fonseca@ufv.br}
\affiliation{Departamento de F\'isica, Universidade Federal de Vi\c cosa, 36570-900,
Vi\c cosa, Brazil}
%%%%%%%%%%%%%%%%%%%%%%%%%%%%

\date{\today}

%%%%%%%%%%%%%%%%%%%%%%%%%%%%
\begin{abstract}
Majorana fermions cannot be found in nature as a free fundamental particle. Nevertheless, in condensed matter systems, they can emerge as a collective excitation. In this work, using functional integration techniques, we calculated the effective potential for emergent Majorana fermions in the Kitaev chain. In this case, we have shown how the superconductor parameter behaves as a function of temperature. Furthermore, we considered surface-induced superconductivity in a Topological Insulator and calculated the effective potential for emergent Majorana fermions in this system. In the case of an s-wave superconductor, we obtained a gap equation equivalent to that one appearing in a quasi-two-dimensional Dirac electronic system, a candidate to explain high-Tc superconductivity. Finally, for the p-wave superconductor, we have obtained a critical value of the electron-electron interaction in the surface of the Topological Insulator, determining the existence or not of induced superconductivity, a remarkable result to guide experiments.

\end{abstract}
%%%%%%%%%%%%%%%%%%%%%%%%%%%%


%\begin{keyword}
%...


\maketitle

%\end{keyword}
%%%%%%%%%%%%%%%%%%%%%%%
%%\end{frontmatter}

% \linenumbers

\section{Introduction}\label{Introduction}

Majorana Fermions \cite{etore, wilczek} are exotic particles that have been studied in high energy physics for decades but have not been observed yet \cite{Elliot}. Proposed in 1937 by Ettore Majorana, they are associated with real solutions of the Dirac equation \cite{etore}, and, consequently, they are their own antiparticles, with an electric charge null. In the frame of field theory, particles like electrons are described by their energy, momentum, and spin. In a solid, an electron is considered as a particle occupying an energy level, while a hole describes an unoccupied level. Majorana fermions can emerge as a quantum superposition of an electron and as a hole that moves freely, each one of them having the same direction or spin. The spin of the Majorana fermion can interact with the spin of atomic nuclei in the material, and therefore, it can be detected by using nuclear magnetic resonance techniques \cite{Elliot}.

From a high energy physics perspective, Majorana fermions are essentially a {\it half} of an ordinary Dirac fermion. Due to the particle-hole redundancy, a single fermionic state is associated with each pair of $\pm E$ energy levels, while the presence or absence of a fermion in this state defines a two-level system with energy splitting $E$. The existence of the Majorana fermions may be helpful to explain why the universe has a final asymmetry between matter and antimatter, once Majorana neutrinos obey all of the Sakharov requirements \cite{Sakharov}. However, from the perspective of solid-state Physics,  if Majorana fermions emerge spatially separated as a zero-mode state \cite{Mourik}, they could be used to encode quantum information \cite{Nayak,Litinski}, and also to tune the heat and charge transport \cite{Ricco}.
 
The formalism of second quantization can be used to describe many-body electronic states. In this framework, electrons are represented by creation and annihilation operators. Under these assumptions, each Majorana fermion of the system can be given by a superposition of particle and antiparticle as

%Each electron of the system can be seen 
%as a superposition of two Majorana fermions as
%
%
\begin{eqnarray}\label{gammac}
\gamma_{j1} = c_{j}^{\dagger} + c_{j},
\quad
\gamma_{j2} = i(c_{j}^{\dagger} - c_{j}),
\end{eqnarray}
where $ c_{j}^{\dagger} $ and $ c_{j} $ are respectively the creation and annihilation operators for electrons, whose quantum numbers are  denoted by the index $ j $. Additionally, $ \gamma $ annihilates or creates a Majorana fermion. Because $ \gamma^{\dagger} = \gamma $, the creation or destruction of such particle produces the same effect into the system. 
Once Majorana fermions consist of a superposition of electron and hole degrees of freedom, we look for materials with this property. Indeed, systems with superconductor order can exhibit such kind of collective behavior, where their quasiparticles are a product of such superposition \cite{BCS}. Hence, Majorana fermions are expected to emerge in superconductor materials. However, in most physical systems, the two Majorana fermions comprising the electron are interlaced, and thus, to describe them as isolated particles make no sense. 

Spatially separated Majorana Fermions can be obtained even when the system presents topological properties, e.g., the Kitaev chain \cite{Kitaev}, a topological insulator (TI) with induced superconductivity on the surface \cite{FuKane}, topological superconductors \cite{Bernevig}, and others. In this way, it is an important issue to obtain and understand the effective potential of emergent Majorana fermions in different contexts. 
The knowledge of this effective potential could be used, for example, to obtain thermodynamic properties, to show how temperature interferes in the topological phase, 
and to perform computational simulations of these quasiparticles in the presence of interactions,
 as well as to describe other physical properties of condensed matter systems in which such particle-like structures appear.
 Finally, determining the effective potential of Majorana fermions can also be useful in optimizing and controlling of topological computation.

In this work, we obtain the effective potential for emergent Majorana fermions on both, the one-dimensional Kitaev Model (KM) and a TI surface with induced superconductivity. The analytical calculations have been performed by using the functional integral techniques in the complex time representation, extracting the natural logarithm of the partition function \cite{Kapusta, Schad}. In the case of KM, the calculation of the effective potential allows us to determine the dependence on the temperature of the superconductor gap. 
Besides, by analyzing the mixture of electrons and holes in the system, we show the existence of a phase transition from trivial to topological. Such results motivate us to use similar techniques to find the effective potential for a TI surface with induced superconductivity \cite{Elliot, Bernevig}. It is shown that this effective potential depends on the type of superconductor gap. For $ s $-wave, the obtained result agrees with high-T$_{c}$ superconductivity theory for Cuprates, which assume that superconductivity appears in the $ CuO $ planes. This exciting result provides some insights into the physical mechanism behind the superconductivity in high-T$_{c}$. For $ p $-wave, the obtained effective potential at zero temperature shows the existence of a continuous quantum phase transition separating the trivial and the superconductor states as a function of the electron-electron interaction. For non-zero temperature, the gap equation also exhibits the qualitative behavior of the superconductor parameter as a function of the temperature.

This work is divided as follows: In section \ref{Kitaev}, we present the results and discussion for the effective potential of a KM. Section \ref{SC2D} brings the discussions on the effective potential of Majorana fermions appearing in IT with induced superconductivity. Finally, in Section \ref{Conclusions}, we present our conclusions and perspectives on the obtained results.

%%%%%%%%%%%%%%%%%%%%%%%%%%%%
\section{The Kitaev Model}\label{Kitaev}

The Kitaev chain is the simplest model exhibiting Majorana fermions. It consists of a chain (1D system) of spinless fermions that interacts with the nearest neighbor. It is an exactly soluble model and provides a useful place to study Majorana fermions in 1D space \cite{Elliot}. This model describes the most straightforward possibility for a 1D superconductivity order with spinless fermions. The Hamiltonian of KM 
can be written as
\begin{eqnarray} \label{Hfermion} 
 \mathcal{H} = \sum_{j} \left[-t\left( c_{j}^{\dagger} c_{j+1} + c_{j+1}^{\dagger} c_{j} \right)
- \mu \left( c_{j}^{\dagger} c_{j} - \frac{1}{2} \right)
\right.
\nonumber \\
\left.
+\Delta \left( c_{j}^{\dagger} c_{j+1}^{\dagger} + c_{j+1} c_{j} \right) \right]
- \frac{\Delta^{2}}{g},
\label{Hkitaev}
\end{eqnarray}
where $ t $ and $ \Delta $ are, respectively, the hopping and the superconductor parameter, $ \mu $ is the electronic chemical potential, and $ g $ is the electron-electron interaction. 
The last term in Eq. (\ref{Hkitaev})
comes from a mean-field calculation over the quartic term, which describes superconductivity \cite{Tinkham}. 
Assuming that $\Delta$ is real and using the inverse representation of Eq. (\ref{gammac}), given by
\begin{equation}
\label{fermionmajorana}
c_{j} = \frac{1}{2} \left(\gamma_{j1} + i \gamma_{j2} \right), \quad \quad 
c_{j}^{\dagger} = \frac{1}{2} \left( \gamma_{j1} - i \gamma_{j2} \right),
\end{equation}
\noindent
the Hamiltonian of KM can be rewritten in the Majorana basis as
\begin{eqnarray}
\label{MajoranaHamiltonian}
\mathcal{H} 
= 
\frac{i}{2} \sum_{j}
\left[
-\mu \gamma_{j,1} \gamma_{j,2}
+ \left(
t + \Delta
\right) \gamma_{j,2} \gamma_{j+1,1}
\right.
\nonumber \\
\left.
+ \left(
-t + \Delta
\right) \gamma_{j,1} \gamma_{j+1,2}
\right] 
-\frac{\Delta^{2}}{g}\,.
\end{eqnarray}

The partition function for Majorana fermions can be then obtained by 
performing a functional integral in time complex representation \cite{Kapusta}, that is,
\begin{equation}
\mathcal{Z} = 
\int \left[ i d\gamma^{\dagger} \right]
\left[ d \gamma \right]
\exp \left[ -\int_{0}^{\beta} d \tau \sum_{j} \left(
\gamma^{\dagger}_{j} \partial_{\tau} \gamma_{j} + \mathcal{H} \right)\right]
\label{partitionmajorana1},
\end{equation}
where $\beta = 1/T$, and $ \mathcal{H} $ is a $2\times2$ matrix
that depends on the  Nambu fields $ \gamma^{\dagger}_{j} = \left( \gamma_{j,1}^{\dagger}\,\,\,\gamma_{j,2}^{\dagger} \right) $ and 
$ \gamma = \left( \gamma_{j,1}\,\,\,\gamma_{j,2} \right)^{T} $,
in the real space. It is worth noticing that those fields are independent, and they can be integrated separately. To calculate this integral, we can proceed with a Fourier transformation of the Majorana operator, in order to obtain a partition function in the momentum space. 
That is,
\begin{eqnarray}
\label{partitionmajorana}
\mathcal{Z} = \int \left[ i d\gamma^{\dagger} \right]
\left[ d \gamma \right]
\exp \left[ -\sum_{n,p} \beta \left( i\gamma_{n,p}^{\dagger} \mathcal{A} \gamma_{n,p} + \frac{\Delta^{2}}{g}\right)
\right],
\end{eqnarray}
where,
\begin{eqnarray}
& 
\gamma_{n,p}^{\dagger} = \left( \gamma_{1,n,p}^{\dagger}\, \,\gamma_{2,n,p}^{\dagger} \right);
\quad
\mathcal{A} =
\left( \begin{array}{cc}
 \omega_{n} &  D^{*} \\
 -D & \omega_{n} \\
\end{array} \right);
\nonumber \\
&
D = \left[\mu + 2 t \cos \left( p \right) + 2 i \Delta \sin \left( p \right)\right]/4,
\end{eqnarray}
$ \omega_{n} $ are the Matsubara frequencies for fermions that can be used in this case \cite{Kapusta, Maldacena}.
 
Eq. (\ref{partitionmajorana1}) and (\ref{partitionmajorana}) are not merely the electronic partition functions of the Hamiltonian (\ref{Hfermion}), with a transformation (\ref{fermionmajorana}) implemented. They consist of a Majorana partition function since the integration is performed over the Majorana fields $\gamma$, as in (\ref{partitionmajorana1}) and (\ref{partitionmajorana}), unlike the partition function for electrons, which integrates over the electronic fields $c$. A simple change of basis in the complete electronic partition function does not provide the correct answer to the partition function for Majorana fermions, because the integration measure for electrons in the Majorana basis is different from the Majorana one used in equation (\ref{partitionmajorana1}).
Indeed, Eq. (\ref{partitionmajorana1}) is the Majorana partition function obtained from Hamiltonian (\ref{MajoranaHamiltonian}) \cite{Maldacena}. This model presents a topological phase when $ |2 t| > |\mu| $, and it is trivial otherwise. In the topological phase, the solutions show spatially separated 
Majorana zero modes (MZM). 

\begin{figure}
[h]
\centerline
{
\includegraphics[width=1.\columnwidth]{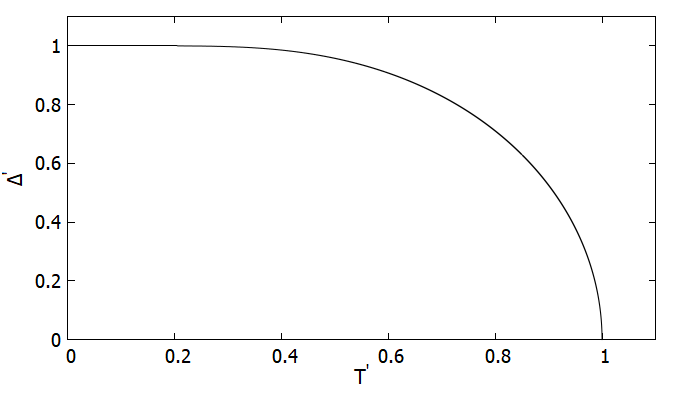}
}
\caption{$ \Delta' $ as function of $ T' $ for the Kitaev Model (normalized values). }
\label{TxDelta}
\end{figure}

In the momenta space 
\begin{eqnarray}
\left\{
\gamma_{i,n,p}^{\dagger}, \gamma_{j,n',p'}
\right\}
= 2 \delta_{i,j} \delta_{n,n'} \delta_{p,p'}.
\label{Grassmann}
\end{eqnarray}
From the property that $ \gamma_{i,n,p}^{\dagger} = \gamma_{i,-n,-p} $,
and assuming $ p = -p' $ and $ n = -n' $,
Eq. (\ref{Grassmann}) reduces to
$ \left\{
\gamma_{i}, \gamma_{j}
\right\} = 0 \Rightarrow \gamma^{2} = 0 $,
which implies that Majorana operators work as Grassmann variables
in momenta space
\cite{Schad}.
In this way, the integral given in Eq. 
(\ref{partitionmajorana}) is reduced to a Gaussian integral, whose effective potential can be promptly 
calculated. Therefore, after performing the sum on the Matsubara frequencies, we have that the effective 
potential is given by
\begin{eqnarray}
\label{Veff1}
\text{V}_{\text{eff}}  & = & 
-\frac{1}{\beta} \ln \mathcal{Z} 
\nonumber \\
& = & \frac{\Delta^{2}}{g} -\frac{1}{\beta} \int dp \left[
\beta |{D}| + 2 \ln \left(1 + e^{- \beta |{D}|}\right) \right] \,.
\end{eqnarray}
%
%
%It must be highlighted that the sign of the chemical potential distinguishes electrons (negative 
%chemical potential) and holes (positive chemical potential). Nevertheless, unlike calculations performed 
%for free fermions \cite{Kapusta}, as a consequence of the system's topology, the contributions of 
%electrons and holes to the effective potential must not be separated. This fact will be evidenced when 
%we perform the calculation of the number of particles. 
The characteristic behavior of the superconductor parameter as a 
function of temperature can be obtained from using the minimum of the $V_{\rm eff}$ and assuming that $ \Delta $ must not change if we are considering electronic or Majorana effective potential. Thereby, by assuming the topological case of interest
when $ t = \Delta $ and $ \mu = 0 $, we obtain
\begin{equation}
\label{DxTKitaev}
T' = \frac{\Delta'}{\text{arctanh}\left( \Delta' \right)},
\end{equation}
where we have defined $ T \equiv T' \Delta_{0}/4 $ and $ \Delta \equiv \Delta' \Delta_{0}$, where $ \Delta_{0} = \frac{1}{4}\int dp\, \sin^{2}p $ can be interpreted as a finite value that does not contribute qualitatively to the obtained results. It can be determined by assuming the existence of an energy cutoff, $ \Lambda $, considered as the limit of the first Brillouin zone, $ \Lambda = \pi / a $ ($ a $ is the lattice spacing). In this context, we can notice that 
Eq. (\ref{DxTKitaev}) qualitatively agrees with experimental data for the behavior of the superconductor parameter. That is, $ \Delta $ has a finite and positive value when $ T = 0 $, decreasing to zero as $ T $ increases (See Fig. \ref{TxDelta}).%

The competition between the hopping parameter and the chemical potential results in a phase transition between the trivial and topological phases. Once $ \mu $ controls the occupation of electrons and holes in the system, a correlation between their presence and the trivial/topological phase is evident. That is, given a specific hopping parameter, the control of the trivial/topological state can be performed by changing the particle number, from the injection of electrons or holes into the system. Thus, the topological state can also be verified by
measuring the occupation of electrons and holes.
The particle number can be then determined by \cite{Kapusta}
\begin{eqnarray}
N
& = &
\frac{1}{\beta} \frac{\partial}{\partial \mu} \mathcal{Z}
\nonumber \\
& = &
\frac{1}{16} \int dp
\frac{\mu + 2 t \cos(p)}{|{D}|}
\tanh\left(
\frac{\beta |{D}|}{2}
\right).
\label{Ne}
\end{eqnarray}
\begin{figure}
[ht]
\centerline
{
\includegraphics[width=1.\columnwidth]{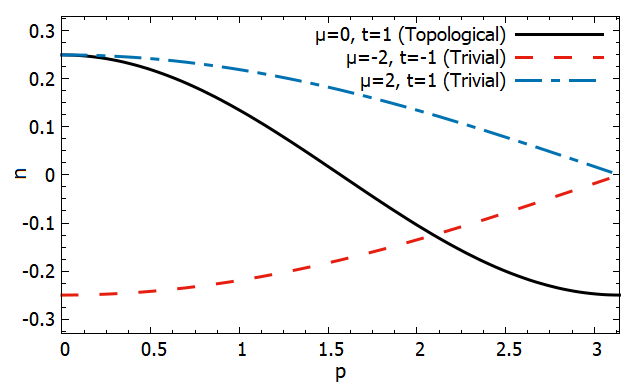}
}
\caption{(Color online) Density particle number in momentum space as a function of $ p $. Here we have adopted 
$ \Delta = 1 $ and different values for $ \mu $ and $ t $.
}
\label{Number}
\end{figure}

The integrand in Eq. (\ref{Ne}) consists of the fermion density in the momenta space, which shows how the topological phase and the presence of MZM occurs, depending on the fermionic density.
That is, if $ |{2 t}| < |{\mu}| $ (trivial phase), the electronic density can be either positive or negative, evidencing that the system has only 
electrons ($\mu > 0$) or holes ($\mu < 0$). On the other hand, 
if $ |{2 t}| > |{\mu}|$, the electronic density signal depends on the momentum, evidencing the existence of a topological phase. In Fig. \ref{Number}, we show the behavior of the number of 
electrons and holes for $ \Delta = 1$ and $T=0$, for different values of $ \mu $ and $ t $. 
In the topological phase, ($ \mu = 0 $ 
and $ t = 1 $), the density of particles changes continuously from positive to negative values. 
Nevertheless, for the trivial phase, the density of particles has the same signal, positive or negative, referring to electrons (blue-dashed line) and holes (red-dashed line). 
This topology is independent of the superconductor parameter, and then, although we have assumed an exact value for $\Delta$, its magnitude is not relevant for the phenomenon, influencing only the amplitude of the particle number. Besides, once temperature effects do not change the topology of the system until the superconductivity disappears, the same effect is observed for non-zero temperature.

\section{Induced Superconductivity on TI Surface}\label{SC2D}

In previous works, Fu and Kane \cite{FuKane} have considered a model where an s-wave superconductor is covering a TI, showing that the superconductivity can be induced on the surface of a 3D TI by proximity effects. Aiming perform a more general analysis, we consider a superconductor with parameter $\Delta(r,r')$, inducing superconductivity in the surface states of 3D TI. The Hamiltonian describing the interface between the superconductor and the TI can be written as
\cite{Elliot,Bernevig,Leijnse}

\begin{eqnarray}
H 
& = & \int dr~dr' \left\{ c_{\uparrow r}^{\dagger} \Delta(r,r') c_{\downarrow r'}^{\dagger}
- c_{\uparrow r} \Delta^{*}(r,r') c_{\downarrow r'} \right. \nonumber \\
& & \left. \hspace{-0.2cm} -\frac{{|\Delta(r,r')|}^{2}}{g}  + \frac{\delta(r - r')}{2} \left[
c_{\downarrow r'}^{\dagger} p_{+} c_{\uparrow r} +c_{\uparrow r} p_{+} c_{\downarrow r'}^{\dagger} \right. \right.
\nonumber \\
& &
\left. \left.\hspace{-0.2cm} +c_{\uparrow r}^{\dagger} p_{-} c_{\downarrow r'} +c_{\downarrow r'} p_{-} c_{\uparrow r}^{\dagger}
- 2 \sum_{\sigma=\uparrow r, \downarrow r' } \mu \left( c_{\sigma}^{\dagger} c_{\sigma} - \frac{1}{2}
\right) \right] \right\}, \nonumber \\
\label{Heletron}
\end{eqnarray}
where $ p_{\pm} = p_{x} \pm i p_{y}$ depends on the momentum operator, $ c_{\sigma}^{\dagger} $ and $ c_{\sigma} $ are the creation and annihilation operators for electrons, and $g$ is the electron-electron coupling, constant in the superconductor. The presence of a vortex in the superconductor parameter can yield zero-mode states with the spinor structure having components that obey the Majorana fermion conditions. It implies in the existence of Majorana bound states in the surface of the TI. Using Eq. 
(\ref{Heletron}), we can determine the effective potential for emergent Majorana fermions using the partition function. For such purposes, the Hamiltonian (\ref{Heletron}) will be rewritten as a function of Majorana operators in the momenta space, using the inverse of relation (\ref{gammac}). Therefore, the partition function written in terms of a functional integral in the time-complex representation has the form
\begin{eqnarray}
\mathcal{Z}
& = &
\left[
\prod_{n} \prod_{p,p'} \prod_{\alpha}
\int 
i d\gamma_{\alpha,n,p,p'}^{\dagger}
d\gamma_{\alpha,n,p,p'}
\right]
\nonumber \\
& &
\exp \left\{ \sum_{p,p'} 
\left[ \sum_{n}  i\gamma_{n,p,p'}^{\dagger}
\mathcal{D}
\gamma_{n,p,p'}
%\right.
%\right.
%\nonumber \\
%& &
%\left.
%\left.
-
\frac{{|\Delta(p,-p')|}^{2}}{g}
\right]
\right\}\,,
\nonumber \\
\label{particaofinal}
\end{eqnarray}
where the Majorana Nambu field is defined as
\begin{eqnarray}
\gamma_{n,p,p'} = \left(
\gamma_{1, \uparrow,n,p} \quad
\gamma_{2, \uparrow,n,p} \quad
\gamma_{1, \downarrow,n,p'} \quad
\gamma_{2, \downarrow,n,p'}
\right)^{T}\,,
\end{eqnarray} 
and the matrix $ \mathcal{D} $ is given by
\begin{eqnarray}
&
\mathcal{D} = \frac{\beta \delta_{p,p'}}{4}
\left(
\begin{array}{cc}
\mathcal{W} + \bar{\mu}  & \mathcal{P}_{-} - \bar{\Delta} \\
\mathcal{P}_{+} + \bar{\Delta}^{*} & \mathcal{W} + \bar{\mu}
\end{array}
\right)\,,
\nonumber
\end{eqnarray}
being
\begin{eqnarray}
&
\mathcal{W} = 
2 \left(
\begin{array}{cc}
\omega_{n} & 0 \\
0 & \omega_{n}
\end{array}
\right),
\bar{\mu} = 
\left(
\begin{array}{cc}
0 & -\mu \\
\mu & 0
\end{array}
\right),
\nonumber \\
&
\mathcal{P}_{\pm} =
\frac{1}{2} 
\left(
\begin{array}{cc}
- i (p_{\pm} + p_{\pm}') & (p_{\pm} + p_{\pm}') \\
- (p_{\pm} + p_{\pm}') & - i (p_{\pm} + p_{\pm}')
\end{array}
\right),
\nonumber \\
&
\bar{\Delta} = 
\frac{1}{\delta_{p,p'}}
\left(
\begin{array}{cc}
i \Delta(p,-p')  & \Delta(p,-p')\\
- \Delta(p,-p') & i \Delta(p,-p')
\end{array}
\right)\,.
\end{eqnarray}
In this set of equations, $ \Delta(p,-p') $ is the Fourier transform of $ \Delta(r,r') $. 
The superconductor parameter can assume different wave configurations, e.g., $ s $-wave or $ p $-wave.  
In both cases, we can write this parameter as 
$ \Delta(p,-p') = i |{\Delta(p)}| \delta_{p,p'} = i |{\Delta}| f(p) \delta_{p,p'} $, 
where $ f(p) $ characterizes the symmetry and $ |{\Delta}| $ is a constant. Also, the integration 
in Eq. (\ref{particaofinal}) can also be solved by using a Gaussian integration. After performing the summation
 over the Matsubara frequencies, the effective potential is (assuming $ \mu = 0 $ for simplicity)
\begin{eqnarray}
\text{V}_{\text{eff}}
& = &
\sum_{p} \left\{
\frac{|{\Delta(p)}|^{2}}{g}
- \sqrt{ |{p}|^{2} + |{\Delta(p)}^{2}|}
\right.
\nonumber \\
& &
\left.
- \frac{2}{\beta} \ln \left[
1 + e^{
- \beta \sqrt{ |{p}|^{2} + |{\Delta(p)}^{2}|}} 
\right]
\right\}.
%\nonumber \\
\label{geralVeff0}
\end{eqnarray}

\subsection{Induced $ s-$wave superconductivity}

Firstly, we will consider the $ s-$wave case, consisting of a thin film of an $s$-wave superconductor inducing superconductivity 
on the surface of a 3D TI.
Therefore, we assume $ \Delta(r,r') = |{\Delta}| \delta(r-r') $, what results in $ f(p) = -i $. The effective potential in the continuum momentum space becomes
\begin{eqnarray}
\text{V}_{\text{eff}}& = &
- 2 \pi \int_{0}^{\Lambda} dp~p \left\{
\frac{2}{\beta} \ln \left[
1 + e^{
+\beta k} 
\right]
- k
\right\}
\nonumber \\
& &
\quad \quad \quad \quad \quad \quad \quad \quad \quad \quad
+ \frac{\pi \Lambda^{2} |{\Delta(p)}|^{2}}{g}\,,
\label{geralVeff}
\end{eqnarray} 
where $ k = \sqrt{ p^{2} + |{\Delta(p)}|^{2}} = \sqrt{ p^{2} + |{\Delta}|^{2}} $. 
To evaluate this integral, we use polar coordinates, whereby the energy cutoff $ \Lambda $  depends on the lattice parameter of the TI. Moreover, we use an approximate circular first 
Brillouin zone (a good approximation for several TI's) and the integration can be performed from using the cutoff as the value of the momentum in the limit of the Brillouin zone \cite{Hasan,Xiao}.
Following the same procedure used in Section \ref{Kitaev}, we can analyze the minimum of the potential in 
order to obtain the behavior of the superconductor parameter as a function of the temperature, which leads to the following 
gap equation
\begin{eqnarray}
\frac{\partial}{\partial \Delta} \text{V}_{\text{eff}}
 = 
- \int_{0}^{\Lambda^{2}} dx~ \frac{2 \pi |{\Delta}|}{\sqrt{|{\Delta}|^{2} + x}}
&&\tanh\left(
\frac{\beta}{2} \sqrt{|{\Delta}|^{2} + x}
\right) \nonumber \\
& &
+ \frac{2 \pi \Lambda^{2} |{\Delta}|}{g}\,.
\label{gap}
\end{eqnarray}

Despite the constants, equation (\ref{gap}) agrees with previous results describing the behavior of $ \Delta $ in quasi-two-dimensional Dirac electronic systems, as a function of the temperature \cite{Marino}. Although described by the effective potential of Majorana fermions, this result was expected since the present work consists of an analysis of an electronic system, with $ \Delta $ being the same parameter in both effective potentials, Majorana fermions, and electrons. Furthermore, it also corresponds to the same results obtained by the spin fermion model, a candidate to explain how high-T$_{c}$ superconductivity emerges in $ CuO $ planes of cuprates systems \cite{Marino2,TeixeiraEPL,TeixeiraSSC}.
The fact that these gap equations are equivalent and both systems are (quasi-)bi-dimensional suggests the existence of a natural emergence of Majorana fermions in the $ CuO $ planes in cuprates. 
Besides that, the superconductor parameter of cuprates has a characteristic form,
which may contribute to the phenomenon as well.
Undeniably, this is an opened issue
and it may be further investigated and corresponds to prospects of continuity of the present work.

\subsection{Induced $ p-$wave superconductivity}

Considering now a $ p-$wave superconductor, we have to insert $ \Delta(p) = |{\Delta}| \left( p_{x} + i p_{y} \right) $ 
in the effective potential (\ref{geralVeff0}),
with $ f(p) = \left( p_{x} + i p_{y} \right) $. 
After performing the Matsubara summation, we obtain the same result 
given in Eq. (\ref{geralVeff}), but with the new $ p $-wave parameter, $\Delta(p) $, instead of the $ s-$wave parameter.

In the system equilibrium, occurring in the minimum of equation (\ref{geralVeff}) (derivative equal zero), 
one can observe the existence of a critical value of the electron coupling constant $ g_c $. Indeed, 
if $g<g_c$, there is no induced superconductivity on the TI surface. Thus, it can be observed that there is a critical $g=g_c$ value allowing induced superconductivity in the surface states of the topological insulator by the proximity effect. This condition leads to the superconductor parameter at zero temperature, $ \Delta_{0} $, as
\begin{eqnarray}
|\Delta_{0}|
= \left\{ \begin{array}{ccc} 0 &,\, ~\text{if}~ & g \leq 4 \Lambda /3 \\
\sqrt{\left( \frac{3 g}{4 \Lambda} \right)^{2} - 1} &,\, ~\text{if}~ & g >  4 \Lambda /3. \\
\end{array}
\right.
\end{eqnarray}
Then, a continuous quantum phase transition occurs in the critical value $ g_{c} = 4 \Lambda /3$, which separates the trivial and the superconductor phases. Additionally, the second derivative is positive, 
evidencing that it is genuinely a minimum of the system.

Aiming at obtaining the magnitude of $ g_{c} $, we must multiply $ p_{\pm} $ in Eq. (\ref{Heletron}) by 
$ \hbar v_{f} $, where $ v_{f} $ corresponds to the Fermi velocity of the surface carriers of the topological 
insulator. In this way, we obtain that $ g_{c} = 4 \hbar v_{f} \Lambda/3 $, evidencing that $ g_{c} $ 
depends only on the TI parameters, once $ \Lambda = 1/a $ is related with the first Brillouin zone.

The described result is very interesting since it implies that if $ g_{sc} < g_{c} $, where $ g_{sc} $ is 
the electron-electron interaction in the superconductor material, the effective electron-electron coupling in the superconductor is not strong enough to induce superconductivity on the TI surface,
due to the properties of the TI. For instance, we can consider the parameters of Bi$_{2}$Se${_3}$, $ \hbar v_{f} = 2.87 $ eV\AA~, and the largest lattice parameter $ a = 30.5 $ \AA~ \cite{Xiao, Lattice}. Hence, we obtain 
$ g_{c} = 0.4 $ eV. Moreover, we can calculate $ g_{sc} $ using \cite{N0}
\begin{eqnarray}
g_{sc} \approx - \frac{1}{N(0) \ln \left( \frac{T_{c}}{\Theta_{D}} \right)},
\label{gsc}
\end{eqnarray}
where $ T_{c} $ is the critical temperature for superconductor transition, $ \Theta_{D} $ is the Debye temperature, and $ N(0) $ is the density of states of the superconductor. If we consider, for example, the Pb parameters, that is, $ T_{c} =  7.19 $ K, $ \Theta_{D} = 105 $ K and $ N(0) = 0.49 $ states/eV \cite{SCparameters,N0}, we obtain $ g_{sc} \approx 0.76 $ eV. Since $ g_{sc} > g_{c} $, it is expected that Pb induces superconductivity in Bi$_{2}$Se${_3}$. As a matter of fact, this result is corroborated by experimental findings in which induced superconductivity has been observed in Bi$_{2}$Se${_3}$ 
in the presence of Pb \cite{SCInduced}. 

At finite temperature, the minimum of the effective potential for $ p $-wave produces the following gap equation
\begin{equation}
\frac{3 g}{2 \Lambda^{3} \sqrt{ 1 + |{\Delta}}|^{2}} 
 \int_{0}^{\Lambda} dp~p~
 \tanh \left(
 \frac{\beta p}{2} \sqrt{1 + |{\Delta|}^{2}}
 \right)= 1.
\end{equation}

\begin{figure}
\centerline
{\includegraphics[width=1.\columnwidth]{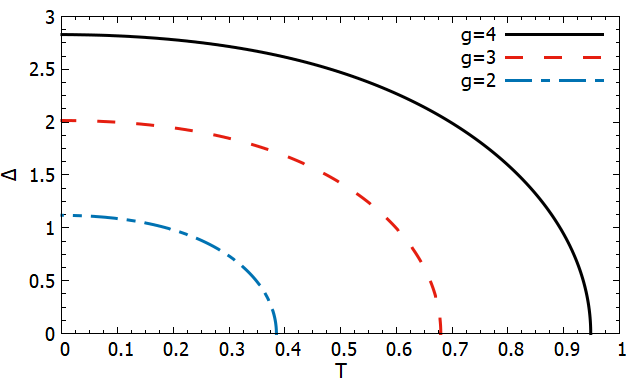}}
\caption{(Color online) $ \Delta $ as function of $ T $ for $ p $-wave superconductor parameter and $ \Lambda = 1 $. 
Note the characteristic behavior of $ \Delta $, which decreases when the temperature increases.}
\label{TxDp}
\end{figure}

The integration of this equation in momenta space results in the dependence of the superconductor parameter on the temperature. 
By assuming $ \Lambda = 1 $, Fig. \ref{TxDp} presents the characteristic behavior of $ \Delta $, which decreases when the temperature increases. Furthermore, it can be noticed that an increase of $ g $ yields an increase of the critical temperature below which the system becomes a superconductor. Also, the value of the superconductor parameter in zero temperature corresponds to the $ |{\Delta_{0}}| $ calculated previously.

\section{Conclusions}
\label{Conclusions}

We have used the functional integral techniques to obtain the effective potential of Majorana fermions for two different systems, which may present such quasiparticle modes spatially separated. These effective potentials 
describe, in both cases, the expected behavior of the superconductor parameter as a function of the temperature.

In the case of the Kitaev chain, we showed that the superconductor parameter has a positive and finite value even when $ T = 0 $, and decreases as $ T $ increases. We have also obtained the density of the particle in momentum space as a function 
of $ p $, showing the differences between topological and trivial phases concerning the predominance of excitations like
 holes and
 electrons into the considered system.
 
From analyzing a topological insulator surface with induced $ s $-wave superconductivity, we have obtained a 
gap equation similar to the one obtained for superconductivity in quasi-two-dimensional Dirac electronic systems. 
Our results suggest the natural emergence of Majorana fermions in $ CuO $ planes. For $ p $-wave and zero temperature, it was showed the existence of a continuous quantum phase transition separating the normal and the superconductor states. The obtained critical electron-electron interaction depends on the parameters of the 
TI. It was also shown that superconductivity would be induced on the surface of IT only if their electron-electron interaction is higher than a critical value of $g_c$. For non-zero temperatures, we have obtained that the gap equation exhibits the expected behavior of the $ \Delta $ as a function of the temperature. Our results are based on systems described by the Hamiltonian presented in Eq. \ref{Heletron}, but they also apply to any system described by equivalent Hamiltonians.

As prospects for future investigations, we will consider the possibility of the emergence of Majorana fermions in the CuO planes in cuprates, which is an intriguing issue to be addressed if we think in applications based on these elusive particles, such as quantum computation.

%%%%%%%%%%%%%%%%%%
\begin{acknowledgments}
This study was financed in part by the Coordena\c c\~ao de Aperfei\c coamento de Pessoal de N\'ivel Superior - Brasil (CAPES) - Finance Code 001. The authors also thank CNPq (Grant Nos. 401132/2016-1 and 309484/2018-9) and FAPEMIG for financial support.

\end{acknowledgments}

\bibliography{apssamp}


%%%%%%%%%%%%%%%%%%%%%%%%%%%%
\end{document}